\documentclass{aastex}

\begin{document}

\title{RADIO/X-RAY LUMINOSITY RELATION\\
       FOR X-RAY BRIGHT GALACTIC NUCLEI:\\
       IMPLICATIONS ON WEIGHING SUPERMASSIVE BLACK HOLE}

\author{Heon-Young Chang$^1$, Chul-Sung Choi$^2$, and Insu Yi$^1$}

\affil{$^1$ Korea Institute for Advanced Study\\ 
207-43 Cheongryangri-dong, Dongdaemun-gu, Seoul 130-012, Korea\\
$^2$ Korea Astronomy Observatory\\ 
61-1 Hwaam, Yusong, Taejon 305-348, Korea}

\email{hyc@kias.re.kr, cschoi@kao.re.kr, iyi@kias.re.kr}

\begin{abstract}
Optically thin and geometrically thick 
accretion flows are known to be responsible for the observed radio/X-ray 
luminosity relation of the X-ray bright galactic nuclei.
It has also been suggested that 
supermassive black hole masses can be estimated  from 
measurements of the core radio luminosity and the X-ray luminosity
by using the advection-dominated accretion flow (ADAF) model. 
In this study we increase the number of data available by compiling
the radio/X-ray fluxes and the mass in published literatures,
and compare the observed ratio of the luminosities 
with predictions from various models of optically thin accretion flows. 
Semi-analytically derived relations of the luminosities 
are presented in cases of 
the standard ADAF model and modified ADAF models, in which 
a truncation of inner parts of the flows and winds causing 
a reduction of the infalling matter are included.
We show that the observed relation can be used indeed to estimate
the supermassive black hole mass, provided that properties of 
such accretion flows are known. 
Having investigated sensitivities of the method on
modifications of the 'standard' ADAF model,
we find that a general trend of model predictions 
from the 'standard' ADAF, the truncated ADAF
and the 'windy' ADAF are somewhat indistinguishable.
We also find, however, that the extreme 
case of the windy model  is
inconsistent with currently available observational data, 
unless microphysics parameters are to be substantially changed.
High resolution radio observations, however, are required to avoid 
the contamination of non-disk components, such as, a jet component,
which, otherwise, results in the over-estimated SMBH mass.
\end{abstract}

\keywords{accretion disks -- black hole physics -- galaxies:nuclei -- 
radiation mechanisms:nonthermal -- radio continuum:galaxies -- 
X-rays:galaxies}

\section{INTRODUCTION}

Supermassive black holes (SMBHs) have been considered 
as the most likely power sources of the activity in quasars and
Active Galactic Nuclei  (Lynden-Bell 1969; Rees 1984). 
SMBHs at the centers of all galaxies are now recognized as 
ubiquitous, whose mass  $M_{\rm SMBH}$ is proportional to the spheroidal 
bulge mass of the host galaxy or the galactic bulge luminosity 
(Kormendy \& Richstone 1995; Magorrian et al. 1998; Richstone et al. 1998)
and is strongly correlated with the velocity dispersion of
the host galaxy (Ferrarese \& Merritt 2000; Gebhardt et al. 2000a; 
Ferrarese 2002). Recently, evidence for the existence of a SMBH 
 in the center of our Galaxy has been added 
(Eckart \& Genzel 1997; Genzel et al. 1997; Ghez et al. 1998).
Though searches for SMBHs are primarily based on spatially resolved kinematics,
another possible way to infer the presence of SMBHs and to shed light 
on the physical condition is to examine spectral energy distribution
 over a wide range from the radio to the hard X-ray frequencies. 
This emission spectrum is produced by an accreting matter, 
as the surrounding gas accretes onto the central SMBH 
(e.g., Frank et al. 1992; Ho 1999). 

Several attempts have been made in utilizing this fact 
to estimate SMBH masses (Yi \& Boughn 1998, 1999; Franceschini et al. 1998; 
Lacy et al. 2001; Ho 2002; Alonso-Herrero et al. 2002; Cao \& Jiang 2002).
It has been also suggested that 
advection-dominated accretion flows (ADAFs) are responsible 
for the observed radio and X-ray luminosities of some of the X-ray bright 
galactic nuclei (Fabian \& Rees 1995; Di Matteo \& Fabian 1997;
Yi \& Boughn 1998, 1999). Moreover, Yi \& Boughn  (1999) discussed that 
the observed radio/X-ray luminosity
relation in terms of the 'standard' ADAF model and demonstrated that 
it could be used as a relatively effective and consequently suitable
 tool to estimate unknown black hole masses.
In this study we further explore this possibility both by 
increasing the number of data available and by deriving similar
relations for the models with additional features, such as, 
truncations, winds, and investigate sensitivities of the method 
Yi \& Boughn suggested on
modifications of the 'standard' ADAF model.

This paper begins with model descriptions for the standard ADAF model 
and modified versions of the model in $\S$  2. We derive analytical expressions to 
describe the radio/X-ray  luminosities for various models. 
We present the calculated ratios of the luminosities and compare them 
with the observational data in $\S$ 3. We discuss and conclude in $\S$ 4.

\section{RADIO/X-RAY EMISSION FROM ACCRETING SMBHs}

When a mass accretion rate is below a critical rate, the radiative 
cooling becomes slower than the viscous heating. As a result of this, 
the dissipated accretion energy is inefficiently radiated away 
and  advected inward to the central object with 
the accreted matter, which is a key idea of ADAF models 
(see Rees et al. 1982; Narayan \& Yi 1994, 1995a,b).
Detailed numerical calculations based on this idea have been performed, 
and the resulting spectra have been successfully applied to a 
number of extra galactic systems 
 (e.g., Narayan et al. 1995; Lasota et al. 1996; 
Mahadevan et al. 1996; Manmoto et al. 1997; Narayan et al. 1998). 

In this study we apply analytical scaling laws for self-similar 
solutions to derive the luminosity relation for ADAF-variations 
(Narayan \& Yi 1994; Mahadevan 1997). We adopt the following 
dimensionless variables : mass of the SMBH $m=M/M_\odot$, where $M_\odot$
is the solar mass; 
radius from the SMBH $r=R/R_g$, where $R_g=2GM/c^2=2.95 
\times 10^5~ m~{\rm cm}$; 
and mass accretion rate $\dot{m}=\dot{M}/\dot{M}_{\rm Edd}$,
where $\dot{M}_{\rm Edd}=L_{\rm Edd}/\eta_{\rm eff} c^2= 
1.39 \times 10^{18}~ m~ {\rm g~ s^{-1}}$ when $\eta_{\rm eff}=0.1$.
We assume that the flows are spherically symmetric as the ADAF is 
quasi-spherical (Narayan \& Yi 1995a). However, the anisotropic extinction of host 
galaxies is not taken into account explicitly. This fact should be 
considered carefully if one wishes to elaborate this method.
The advection fraction $f$ is determined such that the ion and
electron energy balance equations are met. The quantities of 
interest are the volume-integrated quantities which are obtained 
by integrating  the heating rate and cooling rates 
throughout the volume of the flows. For  spherically 
symmetric flows the volume-integrated quantities are defined by
\begin{eqnarray}
Q^X&=&\int^{R_{max}}_{R_{min}} 4 \pi R^2 q^X dR \nonumber \\
&=&3.23 \times 10^{17} m^3 \int^{r_{max}}_{r_{min}}  r^2 q^X dr,
\end{eqnarray}
where $X$ denotes any quantity of interest (cf. Mahadevan 1997). 
As canonical values in a model for the ADAFs parameters are taken to be 
$r_{min}=3$, $r_{max}=10^3$,
$\alpha=0.3$, and $\beta=0.5$, where $r_{min}$ and $r_{max}$ 
represent the inner and the outer boundaries of the disk, respectively,
$\alpha$ is the viscosity parameter, and $\beta$ is
the ratio of the gas pressure to the total pressure. 
The results of this study are insensitive to these values 
unless these parameters are very small.
The lower limit of the integration corresponds to the last stable orbit
for the Schwarzschild black hole, and the upper limit is not very crucial
since most of radiation and heating occur around the central black hole.
We assume that the electron temperature
is constant for $r< 10^3$, as suggested by  Narayan \& Yi (1995b). 

\subsection{Emission from 'Standard' ADAFs}

The total heating by the viscous dissipation for ions
$Q^+$ is given by 
\begin{eqnarray}
Q^+&=&9.39 \times 10^{38} \frac{1-\beta}{f} c_3 m \dot{m} 
r^{-1}_{min}, 
\end{eqnarray}
where $c_3$ is a constant defined as in Mahadevan (1997).
We neglect the possibility of the viscous heating of the electrons, since 
we expect that the fraction of viscous energy transferred to the 
electrons is in the mass ratio of the electron to the ion, $\sim 1/1800$. 
The electrons are assumed to gain energy only by Coulomb interactions 
with the ions. The total heating of the electrons due to 
the Coulomb  interaction is given by   Stepney \& Guilbert (1983)
\begin{eqnarray}
Q^{ie}&=&1.2 \times 10^{38} g(\theta_e)\alpha^{-2} 
c_1^{-2}c_3 \beta m \dot{m}^2 r^{-1}_{min},
\end{eqnarray}
where $g(\theta_e)
=(2+2 \theta_e+\theta_e^{-1})\exp(-\theta_e^{-1})
/K_2(\theta_e^{-1})$, $\theta_e=kT_e/m_ec^2$, $K_2$ 
being a modified Bessel function, 
$k$, $m_e$, and $c$ being the Boltzmann constant,
 the electron mass, and 
the speed of light, respectively,
$c_1$ is a constant given in  Narayan \& Yi (1994).

The synchrotron photons are self-absorbed and give a 
blackbody radiation upto a critical frequency $\nu_c$. 
For a given $T_e$, the synchrotron spectrum 
$L_{\nu}^{sync}$ is given by
\begin{eqnarray}
L_{\nu}^{sync}&=&s_3(s_1s_2)^{8/5}m^{6/5}\dot{m}^{4/5}
T_e^{21/5}\nu^{2/5}, \label{eq:one}
\end{eqnarray}
where $s_1=1.42 \times 10^9 \alpha^{-1/2}(1-\beta)^{1/2}
c_1^{-1/2}c_3^{1/2}$, $s_2=1.19 \times 10^{-13} x_M$, 
and $s_3=1.05  \times 10^{-24}$, $x_M \equiv 2 \nu/3 \nu_b 
\theta_e^2$, $\nu_b \equiv eB/2 \pi m_e c$, 
 $e$ and $B$ being the electron charge,
the magnetic field strength, respectively.  The radio luminosity 
$L_R$ at $\nu$ is defined by $\nu L_{\nu}^{sync}$. The highest 
radio frequency arises from the innermost radius of the accretion 
flows, $r_{min} \sim 3$; $\nu_p=s_1 s_2 m^{-1/2}\dot{m}^{1/2}
T_e^{2} r_{min}^{-5/4}$. At this peak frequency the peak radio 
luminosity is given by
\begin{eqnarray}
L_{R}\equiv \nu_p L_{\nu_p}^{sync}=s_3(s_1s_2)^{3}m^{1/2}
\dot{m}^{3/2}T_e^{7}r^{-7/4}_{min}
 \label{eq:two}.
\end{eqnarray}
We obtain the total power due to synchrotron radiation  from
\begin{eqnarray}
P_{sync}&\simeq&\int^{\nu_p}_{0} L_{\nu}^{sync} d\nu =(5/7) 
\nu_p L_{\nu_{p}}^{sync}.
\end{eqnarray}
We set $\nu_{\rm min}=0$, since  $\nu_{\rm min}$  
is much smaller than $\nu_p$. 

The bremsstrahlung emission per unit volume is given by 
Svensson (1982).
Bremsstrahlung emission is due to both electron-electron 
and electron-ion interactions. 
The total bremsstrahlung power is given by
\begin{eqnarray}
P_{brem}&=&4.74 \times 10^{34} \alpha^{-2} c_1^{-2}
\ln(r_{max}/r_{min})F(\theta_e)m\dot{m}^2, 
\label{eq:eight}
\end{eqnarray}
and the spectrum due to bremsstrahlung emission is given by
\begin{eqnarray}
L^{brem}_{\nu}&=&2.29 \times 10^{24} \alpha^{-2} 
c_1^{-2}\ln(r_{max}/r_{min}) F(\theta_e)m\dot{m}^2
T_e^{-1}\exp(-h \nu/k T_e),  \label{eq:three}
\end{eqnarray}
where 
\begin{eqnarray}
F(\theta_e)&=& 4\Biggl(\frac{2\theta_e}{\pi^3}\Biggr)^{1/2}(1+1.781
\theta_e^{1.34})+1.73\theta_e^{3/2}(1+1.1\theta_e+
\theta_e^2-1.25\theta_e^{5/2}),~~~\theta_e<1,
\nonumber \\
&=&\Biggl(\frac{9\theta_e}{2\pi}\Biggr)[\ln(1.123\theta_e+0.48)+1.5]+
2.3\theta_e(\ln 1.123\theta_e+1.28).~~~~~~~~~~\theta_e>1.
\end{eqnarray}

We neglect the Comptonization of bremsstrahlung 
emission and consider the Comptonization of synchrotron 
emission alone. A contribution by Compton up-scattered 
synchrotron photons to the hard X-ray luminosity becomes 
important as $\dot{m}$ increases, while bremsstrahlung emission 
dominates the hard X-ray luminosity when $\dot{m}$ is substantially low.
We approximate the Comptonized spectrum by assuming that 
all the synchrotron photons to be Comptonized have an 
initial frequency of $\nu_p$. The maximum final frequency 
of the Comptonized photon is $\nu_f=3 k T_e/h$, which 
corresponds to the average energy of the photon for 
saturated Comptonization in the Wien regime. On average, 
we assume that all the photons would see one half the 
total optical depth (Mahadevan 1997), which is written as
$\tau_{es}=6.2 \alpha^{-1} c_1^{-1} \dot{m} r^{-1/2}$.
The spectrum of the emerging photons at frequency $\nu$ 
has the power-law shape
\begin{eqnarray}
L_{\nu}^{Comp} \simeq L_{\nu_i} (\frac{\nu}{\nu_i})^
{-\alpha_c},  \label{eq:four}
\end{eqnarray}
where $L_{\nu_i}$ is the initial luminosity at
 the initial frequency   $\nu_i$ of the photons 
that are Comptonized,  $\alpha_c \equiv -\ln \tau_{es}/\ln A$,
$A$ being the  mean amplification factor in a single scattering.
The total Compton power is therefore given by
\begin{eqnarray}
P_{comp}&=&\int^{3 k T_e/h}_{\nu_p} L_{\nu}^{Comp} 
d\nu \nonumber \\
&=&\frac{\nu_p L_{\nu_{p}}^{sync}}{1-\alpha_c}
\Biggl\{\Biggl[\frac{6.2 \times 10^{10} T_e}{\nu_p}
\Biggr]^{1-\alpha_c}-1 \Biggr\}.
\end{eqnarray}

For  given $m, \dot{m}, \alpha, \beta$, the total heating 
of the electrons should be  balanced
to the sum of the individual cooling; $Q^{ie}=P_{sync}+
P_{brem}+P_{comp}$. The electron 
temperature is obtained by adjusting until this equality is satisfied. 

\subsection{Emission from Truncated ADAFs and 'Windy' ADAFs} 

ADAFs have the positive Bernoulli constant, which is a kind of
total energy of the fluid, and, therefore,
are susceptible to outflows (e.g., Narayan \& Yi 1994).
It might be the case that jets are present near 
the inner region of ADAFs. Radio emission due to ADAFs at the high 
frequency will be suppressed if the inner part of the accretion flow 
is truncated by low frequency jets. In order to simulate this 
situation, we truncate  ADAFs at
a certain inner radius. We suppose that the truncation 
occurs at $\sim 25~ r_g$. An observation of NGC 4486 (M87) 
indicates that the jet is formed in a smaller radius than 
$r \approx 30~ r_g$ (Junor et al. 1999). 
This is also about where is the maximum radius that 15 GHz radio 
emission is generated for $m=10^7, \dot{m}=10^{-3}, T_e=10^9 K$. 
Note that $r$ can be estimated by $r=(1/60)^{-4/5} 
\times (\nu/15~ {\rm GHz})^{-4/5}(m/10^7)^{-2/5} 
(\dot{m}/10^{-3})^{2/5} (T_e/10^9)^{8/5}$.
We consequently set $r_{min}$ in expressions in the previous subsection 
to $25~ r_g$  instead of  $3~ r_g$ in corresponding expressions 
in the truncation case in order to obtain the radio luminosity 
and the X-ray luminosity.
Since the frequency at which the radio luminosity is the highest arises from 
the innermost radius of the accretion disk, $\nu_p \propto  r_{min}^{-5/4}$, the 
truncation radius affects the peak radio luminosity at this frequency according to
Eq. (5).

In addition to a jet or outflows,
since the gas in ADAF solutions is generically unbound
winds may also  carry away infalling matter
(see, e.g., Blandford \& Begelman 1999; Di Matteo 
et al. 1999; Quataert \& Narayan 1999). 
We allow $\dot{m}$ to vary with $r$ as $\dot{m}=
\dot{m}_{out}(r/r_{max})^p$,
where $\dot{m}_{out}$ is the mass accretion rate at $r_{max}$, $p$ is 
the parameter which controls the effects of winds   
(see, e.g., Blandford \& Begelman 1999). The mass loss 
causes a significant effect on a model
spectrum for a large value of $p$. 
Bremsstrahlung and synchrotron emissions decrease with increasing $p$,
due both to the lower density 
 and to the lower $T_e$ in inner parts. Compton emission 
decreases with increasing $p$
even more strongly than the other two emissions since it 
depends both on $\nu_p L_{\nu_p}$ and
$\alpha_c$. 
As $\dot{m}$ is replaced by $\dot{m}_{out}(r/r_{max})^p$, 
expressions for the 'Standard ADAFs'
are to be modified as described below. 
The total heating for ions  $Q^+$ reads
\begin{eqnarray}
Q^+&=&9.39 \times 10^{38} \frac{1-\beta}{f} c_3 m 
\dot{m}_{out}r^{-p}_{max}\frac{1}{p-1}
 (r^{p-1}_{max}-r^{p-1}_{min}),
\end{eqnarray}
where $p<1$. Similarly, $Q^{ie}$ is given by 
\begin{eqnarray}
Q^{ie}&=&1.2 \times 10^{38} g(\theta_e)\alpha^{-2} 
c_1^{-2}c_3 \beta m \dot{m}^2_{out}
r^{-2p}_{max}\frac{1}{2p-1}(r^{2p-1}_{max}-r^{2p-1}_{min})
\end{eqnarray}
where $p \neq 1/2$.
We  assume that  $x_M$ and $T_e$ are constants with $r$
and  $\dot{m}$ as an approximation,
as suggested by Narayan \& Yi (1995b). Given $x_M$, the cutoff 
frequency at each radius is
given as $\nu_c=s_1 s_2 m^{-1/2}\dot{m}^{1/2}_{out}
r_{max}^{-p/2}T_e^{2} r^{-(2p-5)/4}$. 
The synchrotron spectrum is given  by 
\begin{eqnarray}
L_{\nu}^{sync}&=&s_3(s_1s_2)^{\frac{8}{5-2p}}
m^{\frac{4p-6}{2p-5}}
\dot{m}_{out}^{\frac{4}{5-2p}}r_{max}^{\frac{4p}{2p-5}}
T_e^{\frac{21-2p}{5-2p}}\nu^{\frac{4p-2}{2p-5}}. 
\end{eqnarray}
The peak frequency and the radio luminosity at the peak 
frequency are correspondingly
reduced so that we have
\begin{eqnarray}
\nu_p L_{\nu_p}^{sync}=s_3(s_1s_2)^{3}m^{1/2}
\dot{m}_{out}^{3/2}r_{max}^{-\frac{3}{2}p}
T_e^{7}r^{\frac{6p-7}{4}}_{min}.
\end{eqnarray}
The total synchrotron power is also rewritten by
\begin{eqnarray}
P_{sync}&\simeq&\int^{\nu_p}_{0} L_{\nu}^{sync} d\nu 
=\Biggl(\frac{2p-5}{6p-7}\Biggr) \nu_p L_{\nu_{p}}^{sync}.
\end{eqnarray}
For bremsstrahlung emission the total power and the 
spectrum are respectively given by
\begin{eqnarray}
P_{brem}&=&4.74 \times 10^{34} \alpha^{-2} c_1^{-2} 
m \dot{m}_{out}^2 r_{max}^{-2p}
 F(\theta_e)\biggl[\frac{1}{2p}(r^{2p}_{max}-r^{2p}_{min})\biggr], \\
L^{brem}_{\nu}&=&2.29 \times 10^{24} \alpha^{-2} c_1^{-2} 
m \dot{m}_{out}^2 r_{max}^{-2p}
F(\theta_e)T_e^{-1}\exp(-h \nu/k T_e)\biggl[\frac{1}{2p}
(r^{2p}_{max}-r^{2p}_{min})\biggr]. 
\end{eqnarray}
Note that as the exponent $p$ approaches to 0, $[\frac{1}{2p}
(r^{2p}_{max}-r^{2p}_{min})]$ becomes 
$\ln(r_{max}/r_{min})$. For the Comptonization of 
the synchrotron photons we take $\tau_{es}$
as $\tau_{es}=6.2 \alpha^{-1} c_1^{-1} 
\dot{m}_{out}r^{-p}_{max} 
(r^{(2p-1)/2}_{max}-r^{(2p-1)/2}_{min})$.
The Compton spectrum and the total Compton power are 
respectively given by
\begin{eqnarray}
L_{\nu}^{Comp} &\simeq& L_{\nu_i} 
(\frac{\nu}{\nu_i})^{-\alpha_c}, \\
P_{comp}&=&\frac{\nu_p L_{\nu_{p}}^{sync}}{1-\alpha_c}
\Biggl\{\Biggl[\frac{6.2 \times 10^{10} T_e}
{\nu_p}\Biggr]^{1-\alpha_c}-1 \Biggr\}.
\end{eqnarray}

\section{BLACK HOLE MASSES AND RADIO/X-RAY LUMINOSITIES}

Figure 1 shows a plot of the ratio of the 15 GHz  radio 
luminosity to the 2-10 keV X-ray  luminosity versus the 
X-ray luminosity in  logarithmic scales as the luminosities
are derived in the last section. The solid lines 
are standard ADAF model 
predictions for SMBH masses of $10^6-10^9 M_{\odot}$, the 
dotted lines are ADAF model predictions with 
truncations at $25~r_g$, the short dashed lines and 
the long dashed lines are ADAF model predictions with winds, 
where $p=0.4$ and $p=0.99$, respectively.
Values of $p$ considered correspond to the moderate 
wind case, and the strong wind case.
The case of $p=1$ may represent the ADAFs where the convection
is present in ADAFs in that the density scales $-1/2$ with the radius
instead of a power of $-3/2$
(Narayan, Igumenshchev, \&  Abramowicz 2000;  Quataert \& Gruzinov 2000).
For each curve, a constant $\dot{m}$ for 'non-windy' models 
or $\dot{m}_{out}$ for 'windy' models
varies from $10^{-4}$ (upper left corner) to $10^{-1.6}$ (lower
right corner). A general trend of model predictions from 
the 'standard' ADAF, the truncated ADAF, and the 'moderately' windy ADAF 
is somewhat similar and overlapped. 
As long as the canonical  model parameters remain unchanged, 
 however, the 'strongly' windy ADAF model with large $p$, 
which represents the convection,  results in quite different 
trends. Besides the trend, none of data points fall on the area
defined by the convective ADAF model predictions.

The open circles denote the observed core radio 
luminosity at 15 GHz,
and the filled circles the spatially resolved core radio 
luminosity which are converted from ones at
5 GHz to  at 15 GHz using the $\nu^{7/5}$ power law 
(cf. Yi \& Boughn 1999). 
We need such a conversion in order for a direct comparison of
these luminosities at different frequency.
Sources generally fall into the region where theoretical
curves predict. For instance, the mass of the central black 
hole in NGC 3377 is known as about $8 \times 10^7 M_{\odot}$ 
(Kormendy \& Richstone 1995). 
According to our theoretical predictions from the
models except the strong wind ADAF model, the mass
of the central black hole in NGC 3377 falls in between
$10^7 $ to $10^8 M_{\odot}$.
Apparently, even though the inherent radio/X-ray luminosity relation
provides an estimate of the central SMBH masses 
currently available, radio/X-ray flux relations are yet insufficiently 
accurate to give unambiguous estimates of the SMBH mass. Or, 
the observed radio/X-ray fluxes should be analyzed to extract
'disk components' more carefully. 

As for the sources which do not agree with the predictions very well, 
it is thought 
either because they have non-ADAF sources in their centers or because the 
observed radio and/or X-ray flux is due to a non-ADAF origin.
The predicted mass of NGC 224 (M31) is smaller than the estimated mass
at least by two orders of magnitude.
NGC 224 has a very low core radio luminosity, which
is even lower than predicted with the mass accretion rate that 
the X-ray luminosity implies on the basis of an ADAF model. 
NGC 224 seems to harbor a double 
nucleus (Lauer et al. 1993; Bacon et al. 1994).
It is unclear whether ADAF is viable under such circumstance.
Predicted masses of NGC 1068, NGC 3031 (M81), NGC 3079, NGC 4151 are
$\sim 10^8 $ to $\sim$ a few $10^9~ M_{\odot}$. On the other
hand, reported mass estimates of SMBHs of these galaxies are 
less than $\sim 10^7  M_{\odot}$.
Masses of these sources are thought to be over-estimated due to high 
radio/X-ray luminosities resulting from strong jet-related activities.
Moreover, uncertainties in X-ray flux measurements and intrinsic variations
could cause errors in our predictions.
NGC 1068 is famous in that it has a strong compact radio source,
which results in a much higher radio/X-ray luminosity ratio than 
other Seyferts (Gallimore et al. 1996). 
The radio core of NGC 3031 (M81) is also highly variable on many timescale
(Ho 1999).
Moreover, NGC 3031 is substantially brighter in X-rays
relative to the UV than in luminous AGNs (Ho 1999),
and its core luminosity in the 2-10 keV band
varies by a factor of $\sim 2$ in a timescale of years (Ishisaki et al. 1996).
In NGC 3079  the high
radio luminosity may be due to an unresolved small-scale jet.
The radio morphology  of NGC 4151 suggests that the emitting material
has been energized from the nucleus and that the jet interacts with 
the ambient medium (Carral et al. 1990).

In Figure 2, the predicted core radio luminosity is shown as a 
function of the SMBH mass in a logarithmic scale, 
with plots of SMBH candidates whose 
mass estimates are publicly known. From top to bottom, 
for each pair of lines, the mass accretion rate 
corresponds to  $10^{-2}$ and $10^{-4}$. 
The line types  and symbols are same as those in Figure 1.
Although the radio flux data are in high angular resolution
there appears a somewhat wide range of radio luminosities,
which is likely the result of small, and thus invisible, 
radio jets or other components
of various frequencies rather than the pure ADAF emission. 
In particular, the radio luminosity of objects such as NGC 3079, NGC 3031, 
and NGC 1068 seems too high to be explained by the pure ADAF emission.
In Figure 3, the X-ray luminosity for the 2-10 keV band 
is shown as a function of the SMBH mass in a similar way as in Figure 2.
Quoted X-ray fluxes of NGC 4486 and NGC 3377 should be considered as upper limits
(Reynolds et al. 1999; Pellegrini 1999). Note  that the observed
X-ray flux of NGC 3377 in  the energy band of 0.2 - 4 keV is converted to 
the value in the energy band of 2 - 10 keV, assuming 
a power-law with the canonical spectral index 
for Seyfert 1 galaxies, i.e., $\Gamma = 1.7$.
High resolution X-ray images have revealed  discrete X-ray sources for 
NGC 224 (Trinchieri et al. 1999), NGC 3031 (Ishisaki et al. 1996),
NGC 4736 (Roberts et al. 1999). 
For these sources we adopt core or bulge X-ray fluxes.

\section{DISCUSSION}

One of important predictions of the ADAF model 
is the radio/X-ray 
luminosity relation which can be  used to estimate central
SMBH masses. We compute the theoretical predictions of 
the radio/X-ray luminosity  relation
for the ADAF models with  various modifications and compare the predicted
relation with that from the available SMBH candidates.
Several nearby extra galaxies have consistent ratios of radio to X-ray 
luminosities with the ADAF predictions for an estimated mass of the 
central SMBH. If the type of the accretion flow is provided by observations 
with a high angular resolution, 
the inherent radio/X-ray luminosity relation provides a direct 
estimate of the central SMBH mass.
We show the logarithm of the ratio of radio to X-ray luminosities
against the logarithm of the X-ray luminosity for several SMBH masses in Figure 1.
The observations for several extra galactic objects are 
consistent with the predictions of  ADAF models, as
 can be checked in Figures 2 and 3. It is, however, interesting
to note that the model representing the  ADAFs with the strong wind 
can be ruled out unless the microphysics in the standard ADAF model 
has to be modified significantly. 
 
In this study, we implicitly assume that all the SMBHs are non-rotating.
When the black hole rotates, the last stable circular orbit is no longer
3 $r_g$. It becomes smaller than this value and eventually the event
horizon when the black hole spin parameter 
$a \sim 1$. If this is the case the boundary condition
should change as well as the dynamical properties of the flows in 
inner parts (Gammie \& Popham 1998; Popham \& Gammie 1998). Besides dynamics of
the flows, a rotating black hole may power a jet by the Blandford-Znajek 
process (Blandford \& Znajek 1977). 
Therefore, even if the black hole rotates slowly, 
the rotation should be properly taken into account in order 
to make this method robust.
Another crucial modification could be made in terms of microphysics we simplified.
We have assumed that the viscously generated energy mainly heats the ions, 
heat exchanges between ions and electrons occur only due to the Coulomb interaction,
other physical processes, such as, MHD turbulence, pair production, are ignored. Many works
have been done to quantify the effects of such processes (e.g., \"{O}zel et al. 2000).
Once clear understanding is incorporated in the method we present here, its 
implication can be more robust. On the other hand, at the moment, the peak luminosity
of the radio is unlikely to be modified though the 'wings' of the radio spectrum
are modified by the nonthermal electrons. Therefore, as long as we use the peak radio
luminosity instead of whole structure, this method is not seriously defected.

\acknowledgments

We thank the anonymous referee for critical comments which improve
the original version of the manuscript.
IY is supported in part by the KRF grant No. 1998-001-D00365.

\newpage

\figcaption[fig1.eps]{Plot of the ratio of the radio luminosity 
to the 2-10 keV X-ray luminosity versus the X-ray luminosity. 
The solid lines are standard ADAF model 
predictions for SMBH masses of $10^6-10^9 M_{\odot}$, the 
dotted lines are ADAF model predictions with 
truncations at $25~r_g$, the short dashed lines and 
the long dashed lines are ADAF model predictions with winds, 
where $p=0.4$ and $p=0.99$, respectively.
The open circles denote the observed core radio 
luminosity at 15 GHz,
and the filled circles the spatially resolved core radio 
luminosity which are converted from ones at
5 GHz to  at 15 GHz using the $\nu^{7/5}$ power law.
The sources are spatially resolved, and the core luminosities 
are adopted in order to avoid possible contaminations due to
other components such as jets. SMBH masses are denoted by the
log scale at the top of each curve. See the text for detailed discussions.
Radio flux values at 15 GHz of Sgr A*, NGC 1068 are quoted 
from Kormendy \& Richstone (1995) and Gallimore et al. (1996), 
NGC 3079, NGC 3628, NGC 4151, and NGC 4388 are quoted from Carral et al. (1990),
NGC 3031 (M81), NGC 4258, NGC 4736 (M94), and NGC 5194 (M51) 
are quoted from Turner \& Ho (1994), respectively.
Radio flux data at 5 GHz of  NGC 1365 and NGC 3310 are quoted 
from Saikia et al. (1994), 
and NGC 224 (M31), NGC 3377, NGC 4374 (M84), NGC 4486 (M87), 
and NGC 4594 (M104) are quoted from Franceschini et al. (1998), respectively.
X-ray fluxes (2-10 keV) of Sgr A*, NGC 224, NGC 1068, NGC 1365, NGC 3031,
NGC 3079, NGC 3310, NGC 3628, NGC 4151, NGC 4258, NGC 4374, NGC 4388, NGC 4594,
NGC 4736, and NGC 5194 are quoted from 
Baganoff et al. (2001), Trinchieri et al. (1999), Turner et al. (1997),
Iyomoto et al. (1997), Ishisaki et al. (1996), Cagnoni et al. (1998),
Ptak et al. (1999), Ptak et al. (1998), Nandra et al. (1997),
Makishima et al. (1994), Colbert \& Mushotzky (1999), Iwasawa et al. (1997), 
Forster et al. (1999), Ptak et al.(1998), Roberts et al.(1999),
Terashima et al. (1998), respectively. \label{fig1}}

\newpage

\figcaption[fig2.eps]{Core radio luminosity
as a function of the SMBH mass, with SMBHs whose masses are 
available. Mass accretion rates are denoted by the log scale at
 each line. Lines correspond to same models 
as in Figure 1 and symbols represent same observational data points 
as in Figure 1. 
References for radio fluxes are shown in the caption of Figure 1.
The mass estimates of Sgr A*, NGC 1068, NGC 3031 (M81), 
NGC 3079, NGC 4374 (M84), 4594 (M104)  come from 
Eckart \& Genzel (1997), Greenhill et al. (1996), Ho (1999),
Ferrarese \& Merritt (2000), Ho (1999), Ferrarese \& Merritt (2000), 
respectively.
The estimates of SMBH mass in NGC 224 come from 
Dressler \& Richstone (1988), Bacon et al. (1994).
The mass estimate of NGC 3377  comes from
Kormendy \& Richstone (1995) and Magorrian et al. (1998).
The mass of NGC 4151 have been measured by 
reverberation mapping method (Wandel et al. 1999; Gebhardt et al. 2000b).  
The mass estimate of NGC 4258 comes from Miyoshi et al. (1995).
The mass estimates of NGC 4486 (M87) come from 
Ford et al. (1994), Reynolds et al. (1996), Magorrian et al. (1998), and
Ho (1999). 
\label{fig2}}

\newpage

\figcaption[fig3.eps]{Similar plot as Figure 2. 
But the X-ray luminosity for 
the 2-10 keV band is shown with the mass-known SMBHs. 
Lines correspond to same models as in Figure 1.
References for X-ray fluxes are shown in the caption of Figure 1.
References for mass estimates are shown in the caption of Figure 2. \label{fig3}}

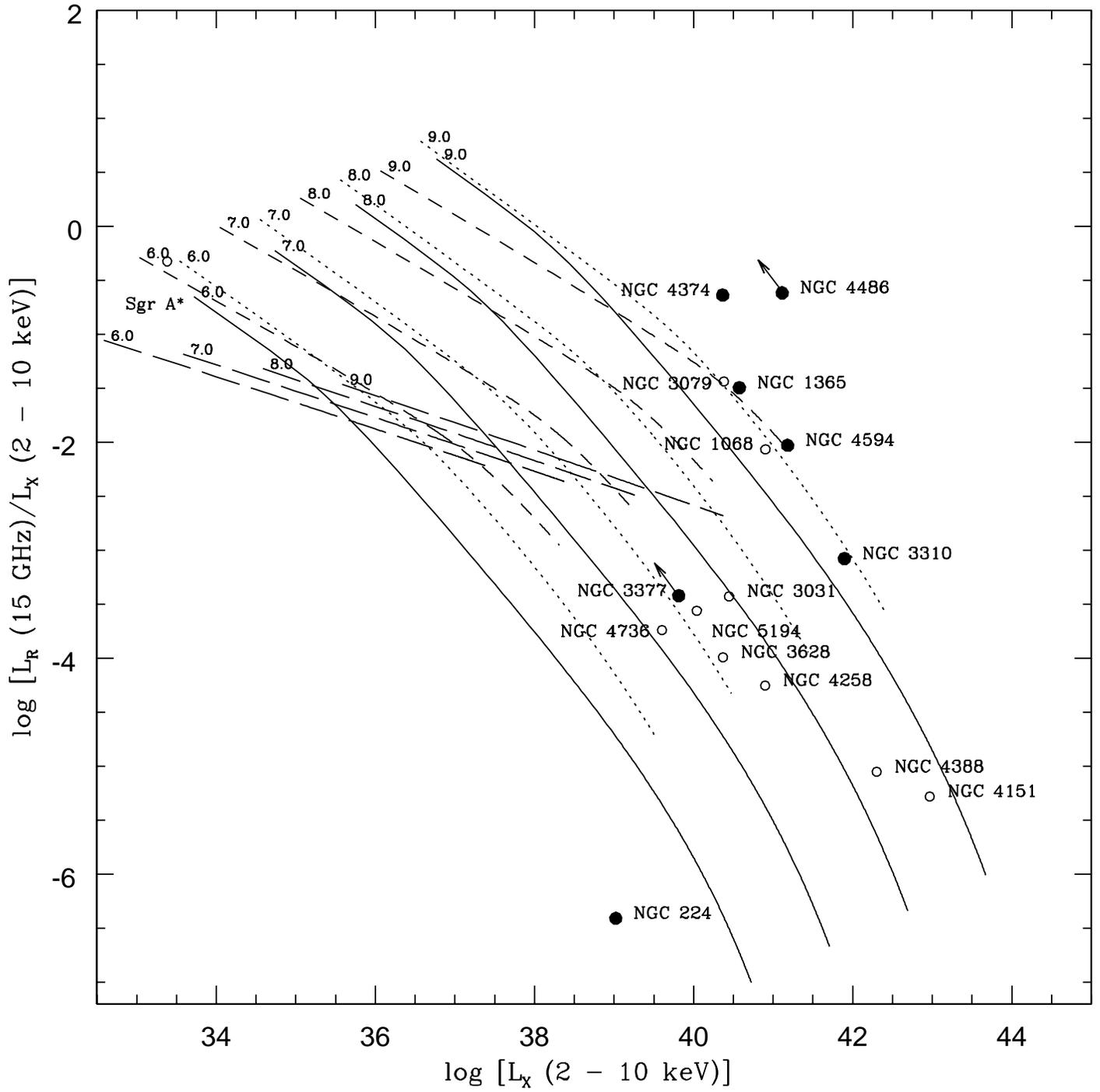
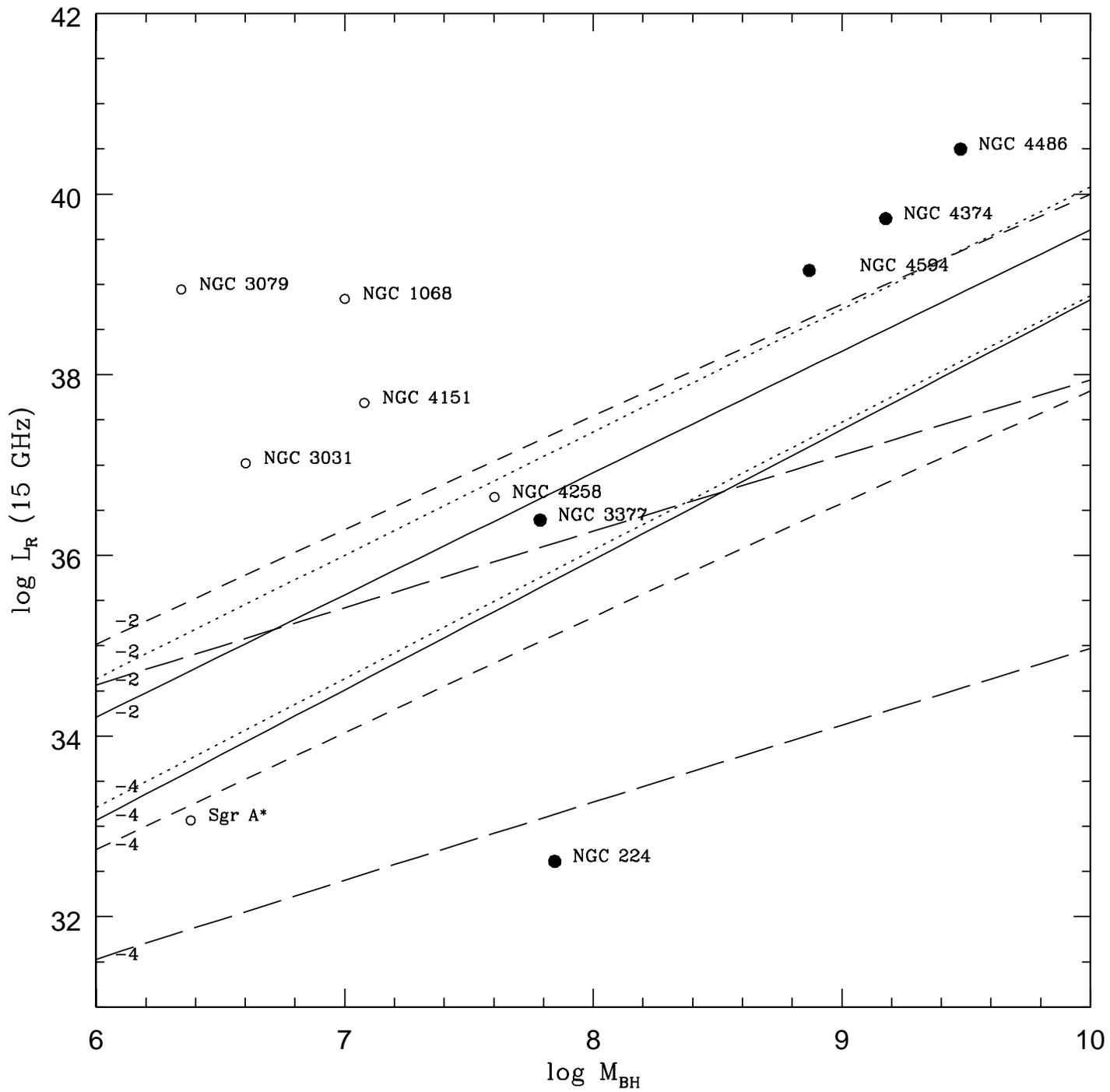
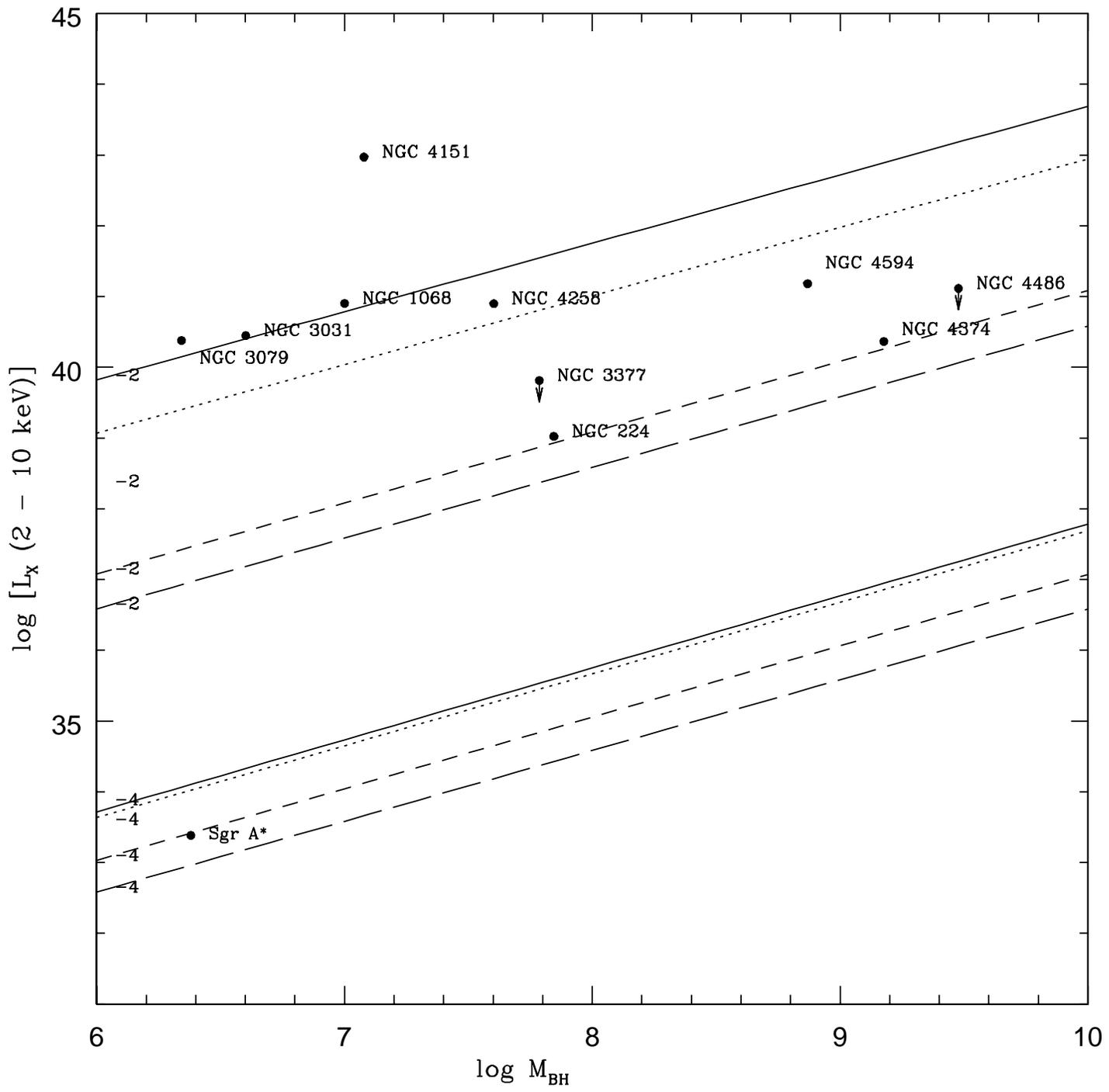
\end{document}